\documentclass[prb,twocolumn, notitlepage,showpacs,preprintnumbers,amsmath,amssymb,10pt,longbibliography]{revtex4-1}

\usepackage{subfigure}

\usepackage{graphicx}
\usepackage{amsmath}
\usepackage{physics}
\usepackage[hidelinks]{hyperref}
\usepackage{amsbsy}
\usepackage{placeins}
\usepackage{comment}

\usepackage{color}
\usepackage{dcolumn}
\usepackage{bm}

\newcommand{\fl}[1]{(#1)} 
\renewcommand\vec[1]{\ensuremath\boldsymbol{#1}}

\newcommand{\kv}{\vec{k}}
\newcommand{\Kv}{\vec{K}}
\newcommand{\rv}{\vec{r}}
\newcommand{\Rv}{\vec{R}}
\newcommand{\tauv}{\vec{\tau}}

\newcommand{\Omegav}{\vec{\Omega}}

\newcommand{\mat}[1]{\mathbf{\underline{#1}}}

\begin{document}

\title{Tunable valley Hall effect in gate-defined graphene superlattices}

\author{Johannes H. J. Martiny}
\email{johmar@dtu.dk}

\author{Kristen Kaasbjerg} 

\author{Antti-Pekka Jauho}

\affiliation{Center for Nanostructured Graphene (CNG), Department of Physics, Technical University of Denmark, DK-2800 Kongens Lyngby, Denmark}

\date{\today}

\begin{abstract}

\noindent 
We theoretically investigate gate-defined graphene superlattices with broken inversion symmetry as a platform for realizing tunable valley dependent transport. Our analysis is motivated by recent experiments [C. Forsythe {\it et al.}, Nat. Nanotechnol. {\bf 13}, 566 (2018)] wherein gate-tunable superlattice potentials have been induced on graphene by nanostructuring a dielectric in the graphene/patterned-dielectric/gate structure.
We demonstrate how the electronic tight-binding structure of the superlattice system resembles a gapped Dirac model with associated valley dependent transport using an unfolding procedure. In this manner we obtain the valley Hall conductivities from the Berry curvature distribution in the superlattice Brillouin zone, and demonstrate the tunability of this conductivity by the superlattice potential. Finally, we calculate the valley Hall angle relating the transverse valley current and longitudinal charge current and demonstrate the robustness of the valley currents against irregularities in the patterned dielectric. 

\end{abstract}

\maketitle 

\section{Introduction}
The electronic structure of graphene hosts well-separated degenerate minima in momentum space which are labeled as the $K, K'$ valleys.\cite{CastroNeto2009} Electrons in graphene are thus described not only by their charge and spin but also by their valley degree of freedom which is conserved when intervalley scattering is absent. In recent years this new degree of freedom has been proposed as a stable carrier of information in so-called valleytronics. \cite{Rycerz2007,Schaibley2016,Yamamoto2015,Mak2014,Vitale2018,Yamamoto2015} 

In hexagonal materials lacking inversion symmetry, control of the valley degree of freedom can be accomplished by generating opposite transverse currents of carriers with different valley index when applying an in-plane electric field. This valley Hall effect is the result of a nonzero Berry curvature of opposite sign in each valley which acts as a valley dependent magnetic field in momentum space. \cite{Xiao2007} 
Indirect measurements of valley currents in such materials have been suggested in e.g. bilayer graphene under transverse electric field,\cite{Shimazaki2015,Li2016,Si2016} or in graphene superlattices defined by an underlying hexagonal boron nitride (hBN) substrate aligned commensurately with the graphene sheet. \cite{Gorbachev2014} These observations have been made in nonlocal transport measurements where a current flowing between two terminals in a Hall bar induces a nonlocal voltage between two different terminals through a combination of the direct and indirect valley Hall effects. 

The valley Hall effect and the associated valley currents are absent in pristine graphene unless perturbations break the sublattice symmetry of the bipartite lattice. The electronic properties of graphene have previously been engineered using e.g. strain,\cite{Si2016,Zhang2017,Settnes2017} substrate effects,\cite{Song2014,Wolf2018,Hu2018} or lithographic etching of a periodic array of holes in the graphene sheet.\cite{Pedersen2008,Sandner2015,Jessen2019} 
Recently, a new approach to band structure engineering has been demonstrated where holes or indentations are made not in the graphene sheet but in an underlying dielectric instead.\cite{Forsythe2018} This procedure avoids introducing any short range disorder to the graphene sheet, and thus limits intervalley scattering while effectively inducing a superlattice potential on the graphene sheet by a gate under the dielectric. As such, this nanostructuring approach seems very well suited for valleytronic applications.   

In this work we theoretically investigate the electronic structure and valley dependent properties of a graphene superlattice geometrically structured for valleytronics. We define a superlattice by a periodic external potential corresponding to a graphene sheet gated through a nanostructured dielectric with a regular array of indentations or holes. Symmetry analysis of this structure reveals that a finite valley Hall effect is possible when these holes do not have an inversion center. Our choice of superlattice structure is supported by earlier studies demonstrating extremely stable band gaps with respect to disorder when perturbations break the graphene A/B sublattice symmetry,\cite{Gregersen2018,Malterre2011} and by the natural formation of such deformations in hBN. \cite{Ryu2015} 

We study the electronic band structure of these systems within a tight-binding model and show the emergence of tunable band gaps in the energy spectrum as the superlattice potential is applied. Using an unfolding procedure for the spectral weight and electronic Berry curvature,\cite{Olsen2015} the superlattice results are mapped to the graphene Brillouin zone where we recover a gapped $K, K'$ valley structure with Berry curvature distributions of opposite sign in each valley. We compare these supercell tight-binding results with an analytical model of graphene with sublattice asymmetry and an overall shift in the Fermi energy, and find a close resemblance at small superlattice potentials. We furthermore compute the valley-resolved transverse conductivities arising from the finite Berry curvature distributions in each valley, and demonstrate the tunability of these conductivities with the strength of the applied superlattice potential, as well as the position of the Fermi energy. Finally, 
a Boltzmann equation approach for the longitudinal conductivity enables us to calculate the valley Hall angle at different electronic fillings and make predictions for experimental observations in nonlocal transport experiments. \cite{Beconcini2016}

\begin{figure}[tb]
\includegraphics[width =  \linewidth]{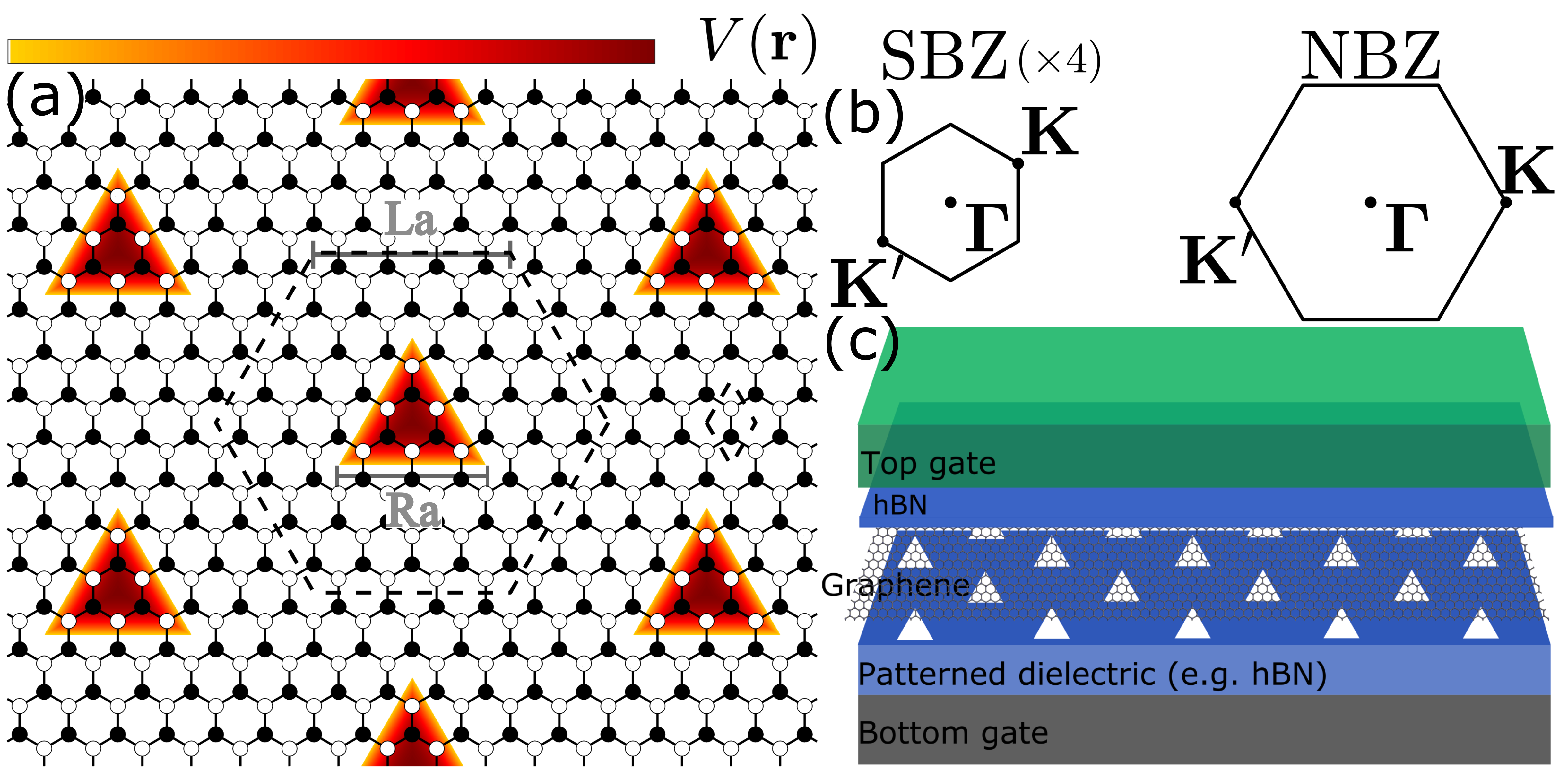}
\caption{
\fl{a} The superlattice system considered in this work: a graphene sheet (empty and filled circles) gated through a patterned dielectric with triangular zigzag-edged holes yielding an effective superlattice potential (red-to-black gradient). The supercell is marked by the dashed lines (left), alongside the normal (graphene) unit cell (right). The lack of inversion center and the sublattice asymmetric structure of the gated regions induce the valley Hall effect under in-plane electric field. The geometry is characterized by the supercell hexagon side length $L$ and the triangle side length $R$. 
\fl{b} The corresponding supercell (SBZ) and normal (NBZ) Brillouin zone. The SBZ is shown enlarged four times for clarity. \fl{c} Sketch of the considered graphene/nanostructured dielectric/gate structure. Here we show nanopatterned hBN with the naturally occurring triangular zigzag edges holes nucleated on boron sites.  
}
\label{fig:intro}
\end{figure}

\section{Method}
We consider a graphene sheet under the effect of a periodic superlattice potential, providing a model for graphene on top of a patterned dielectric. We posit a triangular array of holes etched into the dielectric, and thus a similar structure for the induced superlattice potential in the graphene monolayer as shown in Fig. \ref{fig:intro}. The hexagonal unit cell of this superlattice is shown in Fig. \ref{fig:intro}\fl{a}, with the induced gate potential indicated by the gradient. The geometries considered can be uniquely described by the supercell hexagon side length $L$, and the triangular superlattice potential side length $R$.  
We model the superlattice by a tight-binding Hamiltonian which includes onsite terms arising from the gate induced potential
\begin{align}
H  &= \sum_{i, \sigma} V(\vec{r}_i)c^\dagger_{i\sigma} c_{i\sigma} +  \sum_{\langle ij \rangle, \sigma} t_{ij} c^\dagger_{i\sigma} c_{j\sigma}
\end{align}
where $t_{ij} = -t\delta_{\langle ij \rangle} $, with $t = 3.033$ eV, includes nearest neighbor hopping, and  
$V(\vec{r})$ is the gate-induced potential, defined here along a zigzag edge in the graphene sheet since this edge profile minimizes intervalley scattering.\cite{Brey2006,Orlof2013} The potential corresponds to a zigzag edged triangle etched into e.g. hBN as the dielectric, where such perforations appear naturally nucleated on a single sublattice.\cite{Ryu2015} In our calculations we consider both perfectly sharp (flat) and smoothly varying spatial profiles of the potential, as well as some degree of armchair edges caused by edge disorder in the dielectric nanostructuring. In the following we ignore the possible lattice constant mismatch between the hBN and graphene, and the resulting moir\'{e} structure. Other inversion symmetry breaking shapes of the induced superlattice potential can also lead to the valley Hall effect in the superlattice. Here we restrict ourselves to the $C_3$ structures outlined above, wherein stable band gaps and lack of intervalley scattering lead directly to characteristic plateaus of finite valley Hall conductivity.

Our main goal is to calculate the transverse conductivity arising from the valley Hall effect. This effect can be understood from wave-packet dynamics.\cite{Xiao2010,Culcer2005} The equation of motion for such a wave-packet composed of states from a single band $n$, can in the presence of an electric field be written ($\hbar = 1$)
\begin{align}
\vec{\dot{r}}_n(\kv) &= \partial_{\kv} \epsilon_n(\kv)  - e \vec{E} \times \Omegav_{n}(\kv) 
\end{align}
where we recognize the first term on the right-hand side as the conventional band velocity, while the second term is responsible for various anomalous transport phenomena, determined by the electronic Berry curvature 
\begin{align}
	\Omega_n(\kv) &= \nabla_{\kv} \cross i \bra{u_{n\kv}} \nabla_{\kv} \ket{u_{n\kv}}, \label{eq:BC_def}
\end{align}
written here in terms of the periodic part of the Bloch state, $\ket{u_{n\kv}} = e^{-i\kv \cdot \rv} \ket{\psi_{n\kv}}$. In particular, when an in-plane $E$-field is applied to a perturbed graphene lattice with broken inversion symmetry, electrons in each valley have opposite Berry curvature and thus acquire transverse anomalous velocity components depending on their valley index, leading to the valley Hall effect. 

Valley resolved conductivities follow from the Berry curvature of occupied states by integrating over each valley region separately
\begin{align}
\sigma^{K (K')}_{xy}(E_F) &= -\frac{2e^2}{h} \int_{K (K')} \frac{d^2k}{2\pi} ~\Omega_{xy}(\kv,E_F). \label{eq:VRHC}
\end{align}
Here, the integration region in each case is exactly half the Brillouin zone with the $\Gamma \to M$ symmetry lines as the borders, \cite{Olsen2015} and we have defined the Berry curvature of occupied states 
\begin{align}
\Omega_{xy}(\kv, E_F) &= \sum_{n} f_n(\kv) \Omega_n(\kv),
\end{align}
with $f_n(\kv) = [e^{(E_{n\kv}-E_F)/k_BT}+1]^{-1}$ the Fermi-Dirac distribution. We fix a low temperature of $T = 1$ K in the following in order to clearly distinguish the step in the valley resolved conductivity near the band edges. 

The valley Hall conductivity is then defined as the difference between the valley-resolved conductivities
\begin{align}
\sigma^{v}_{xy} = \sigma_{xy}^K - \sigma_{xy}^{K'}. \label{eq:VHC} 
\end{align}
 In the presence of time-reversal symmetry only half the Brillouin zone needs to be considered in the calculation of the valley Hall conductivity since $\sigma_{xy}^K  = - \sigma_{xy}^{K'}$ and thus $\sigma^{v}_{xy} = 2 \sigma_{xy}^K = - 2 \sigma_{xy}^{K'} $. \cite{Xiao2010} 

\subsection{Unfolding}
We now turn to the calculation of the valley-resolved conductivities from the tight-binding supercell results. 
Diagonalization of the tight-binding Hamiltonian yields the supercell eigenenergies and Bloch states $E_{n \kv}, \ket{\psi_{n\kv}}$, from which we can also obtain the spectral function 
\begin{align}
	A(\kv,\omega) &= \sum_{n\kv} \frac{\eta / \pi }{(\omega-E_{n\kv})^2+\eta^2},
\end{align} 
where $\eta$ is a numerical broadening. 

The valley-resolved conductivities are not immediately available since the Berry curvature folds into the superlattice Brillouin zone (SBZ) in a nontrivial way, which prohibits the direct application of Eq. (\ref{eq:VRHC}). Our approach is thus to unfold the Berry curvature obtained in the SBZ back into the graphene (normal) Brillouin zone (NBZ) and recover information about the valley degree of freedom. \cite{Ku2010} We note that the considered superlattice potential is a perturbation clearly described in terms of the underlying ordered graphene lattice, and that the unfolded Berry curvature and associated valley Hall conductivity thus remain well-defined.\cite{Bianco2014,Olsen2015} Details of this unfolding procedure can be found in Appendix \ref{app:UNF}, and we provide here a short summary. 

The central quantity in the unfolding procedure is the overlap between a normal cell orbital $\ket{\chi_{i\kv}}$ with $\kv \in $ NBZ and a supercell Bloch state $\ket{\psi_{N\Kv}}$ with $\Kv \in$ SBZ,   
\begin{align}
	\lambda_{iN\kv} &=  \braket{\chi_{i\kv}}{\psi_{N\Kv}},
\end{align}
which we can calculate directly from the tight-binding Bloch states. 

Quantities in the SBZ can then be unfolded to the NBZ by convolution with the overlap $\lambda$, and, e.g., the unfolded spectral function becomes
\begin{align}
	A^{(u)}(\kv,\omega) &= \sum_i \sum_{N \Kv} \abs{\lambda_{iN\kv}}^2  \frac{\eta / \pi }{(\omega-E_{N\Kv})^2+\eta^2} \label{eq:A_UNF}. 
\end{align} 
where the sum over $i = A, B$ spans the sublattices of graphene, and $E_{N\Kv}$ are the band energies of the superlattice. 
The unfolding of the Berry curvature [Eq. (\ref{eq:BC_def})] from the tight-binding result follows in a similar manner but requires a more extensive treatment, since the analogous expression to Eq. (\ref{eq:A_UNF}) becomes gauge dependent. \cite{Bianco2014,Olsen2015}  
Once the unfolded Berry curvature $\Omega^{(u)}(\kv,E_F)$ is obtained by this procedure, the valley-resolved conductivities follow by a simple application of Eq. (\ref{eq:VRHC}).

\subsection{Valley Hall angle}
We characterize the relative magnitude of the response associated with the valley Hall effect by calculating the valley Hall angle 
\begin{align}
\tan \theta_v &= \frac{\sigma_{xy}^v}{\sigma_{xx}}. 
\end{align}
This angle is finite only close to the band edges where the valley Hall conductivity is nonzero.
We obtain the longitudinal conductivity $\sigma_{xx}$ from a DC Boltzmann equation approach in the relaxation time approximation \cite{Smith1989}  
\begin{align}
\sigma_{xx} &= 2e^2 \frac{1}{A} \sum_{n\kv} \tau_{n\kv} v_{n \kv,x}^2 \delta(E_F - E_{n\kv}) \label{eq:implemented}
\end{align}
where $A$ is the sample area, and  $\vec{v}_{n \kv} = (1/\hbar) \nabla_{\kv} \epsilon_{n\kv}$ is the band velocity component in the $\hat{x}$ direction. Here, we calculate this analytically from the tight-binding Hamiltonian. 
\begin{align}
\vec{v}_{n \kv} &= \frac{1}{\hbar} \mel{n\kv}{\nabla_{\kv} H_{\kv}}{n\kv}. 
\end{align}
For numerical evaluation of the longitudinal conductivity at low temperatures we approximate the delta function by a Lorentzian $\delta(E_F - E_{n\kv}) \to \frac{1}{\pi}(\eta/2)[(E_F - E_{n\kv})^2 + (\eta/2)^2]^{-1}$ with a constant broadening $\eta = 3$ meV. 

We extract the relaxation time from a typical mobility near the charge neutrality point in hBN encapsulated graphene $\mu \approx 10^5$ cm$^2$ V$^{-1}$  s$^{-1}$ at the given temperature. If we consider the conduction to be limited by charged impurities, the relaxation time varies linearly with the Fermi energy \cite{Hwang2007}
\begin{align}
\tau_{k_F} &= C_{ci,\tau} E_F,
\end{align}
where the proportionality constant is $C_{ci,\tau} \approx 10$ ps/eV at the chosen mobility. For gapped systems we set $\tau_{n\kv} = C_{ci,\tau} \delta E_{n\kv}$ in Eq. (\ref{eq:implemented}) where $\delta E_{n\kv}$ is the energy measured from the band edge of the gapped region.

\section{Results}
\subsection{Band structure and Berry curvature in the supercell}
\begin{figure}[tb]
\includegraphics[width = \linewidth ]{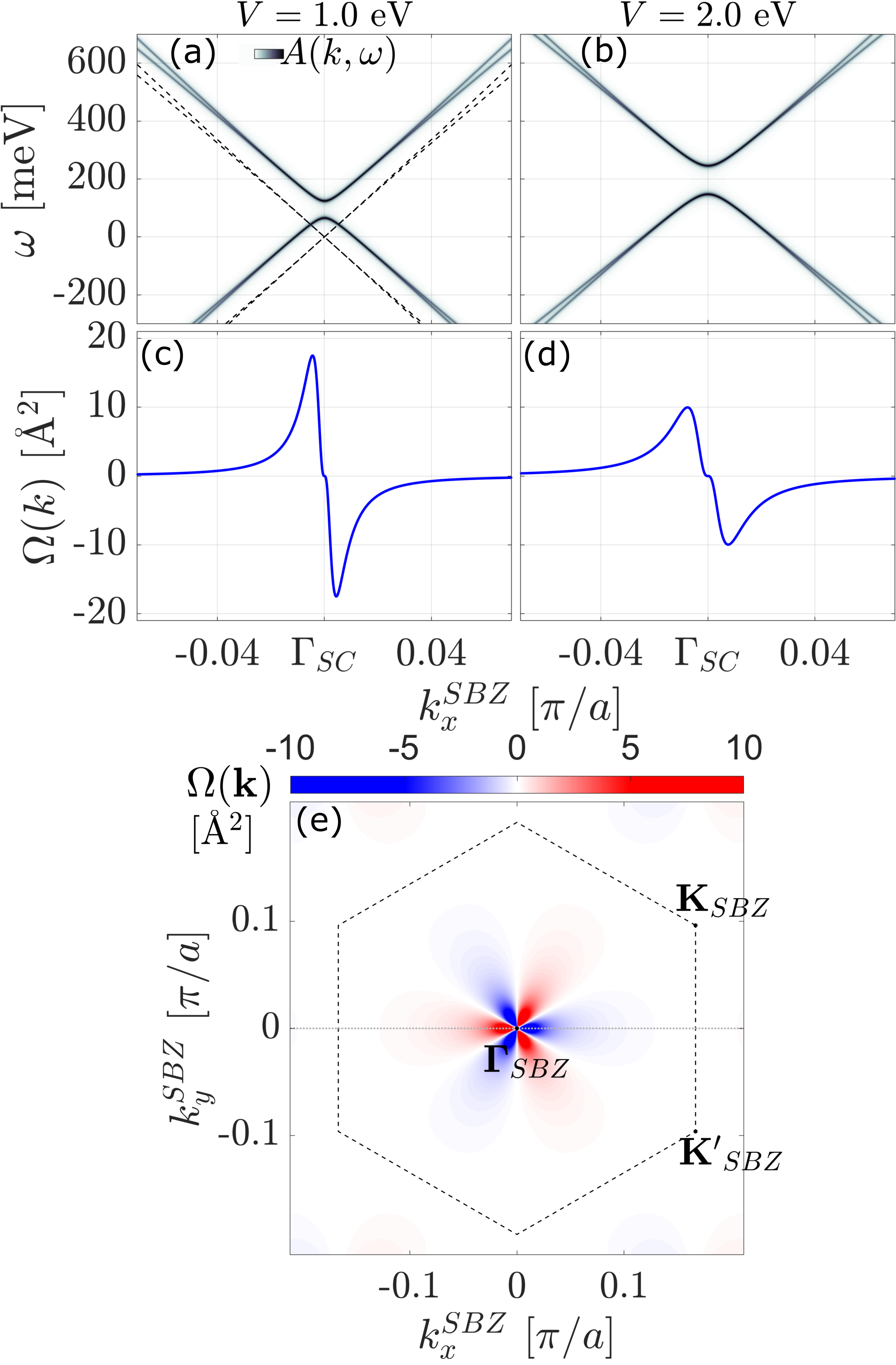}

\caption{
\fl{a}-\fl{b} Spectral weight (gray surface, $\eta = 3$meV) close to the SBZ $\Gamma$ point for different values of the constant superlattice potential $V(\rv_i) = 1$ eV, $2$ eV. The dashed lines in \fl{a} show the $V=0$ (pristine graphene) band structure. 
\fl{c}-\fl{d} Corresponding line-cuts of the occupied Berry curvature when the Fermi energy is fixed in the gap at each potential. 
\fl{e} Supercell Berry curvature in the SBZ with the valence band filled. The pristine system $K$ and $K'$ valleys fold to the SBZ $\Gamma$ point, yielding a sign changing peak centered on this symmetry point. The horizontal dotted line indicates the cut in $\kv$-space shown above. 
}
\label{fig:linecuts_supercell}
\end{figure}

\begin{figure}[tb]
\includegraphics[width = \linewidth ]{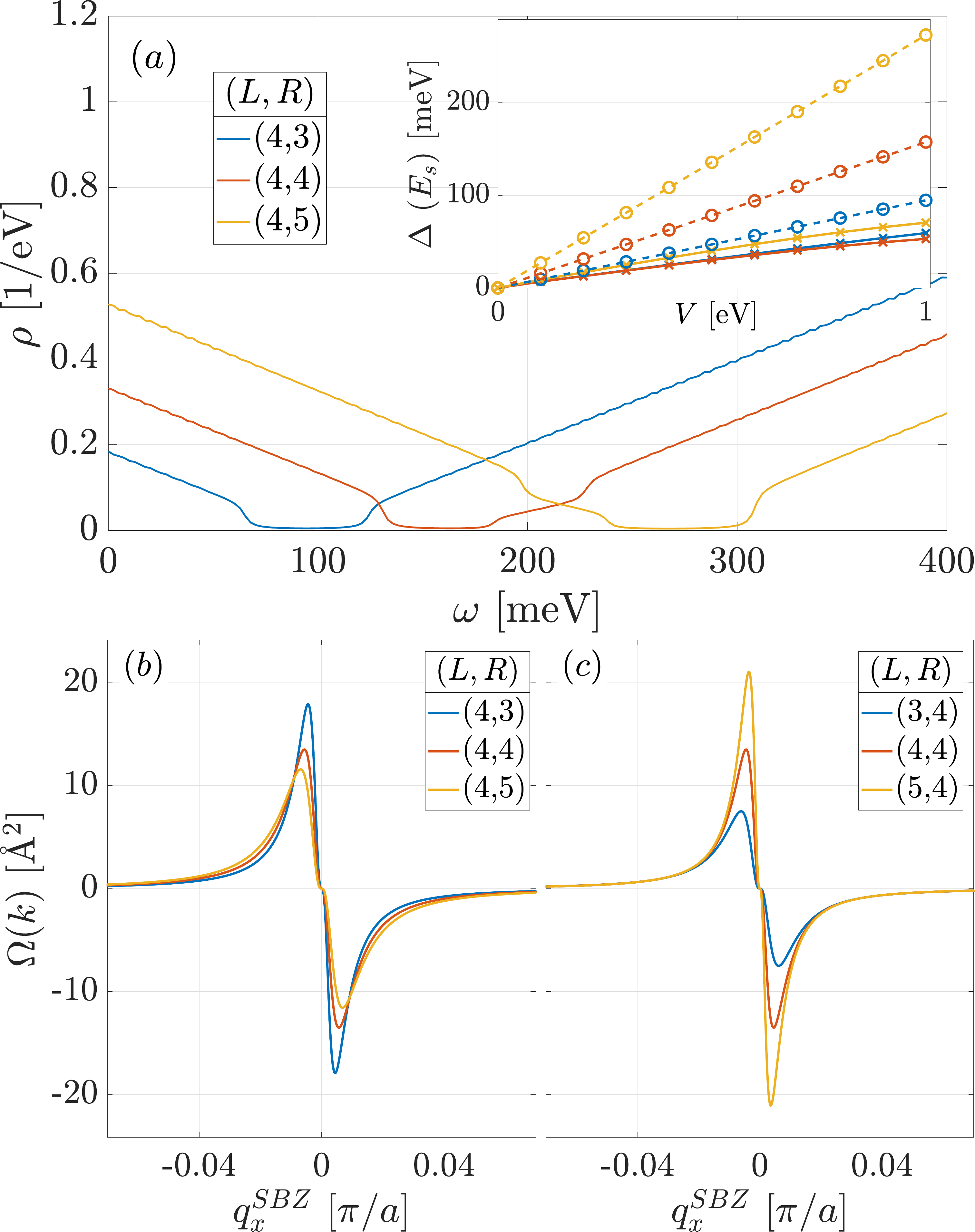} 
\caption{
 \fl{a} Density of states for different geometries of the superlattice at $V= 1$ eV. Inset: Band gaps as a function of the superlattice potential magnitude (full lines) for different geometries of the gated region ($L = 4$). The band gap widens as the superlattice potential is increased in all considered geometries. Dashed lines show the corresponding shift ($E_s$) of the center of the band gap as the superlattice potential is increased. This shift increases linearly with increasing superlattice potential, with the slope determined by the size of the gated region ($R$). 
 \fl{b} Linecut of the SBZ Berry curvature in the gap for different geometries of the gated regions  ($L=4$, $V=1$ eV). The shape of the Berry curvature distribution broadens for increasing size of the gated region $R$, mirroring the broadening with increasing superlattice potential magnitude $V$. \fl{c} The SBZ Berry curvature when the supercell size is varied instead, demonstrating the opposite scaling in the width. 
 }
\label{fig:gap_shift_4}
\end{figure}

\begin{figure}[tb]
\includegraphics[width = \linewidth ]{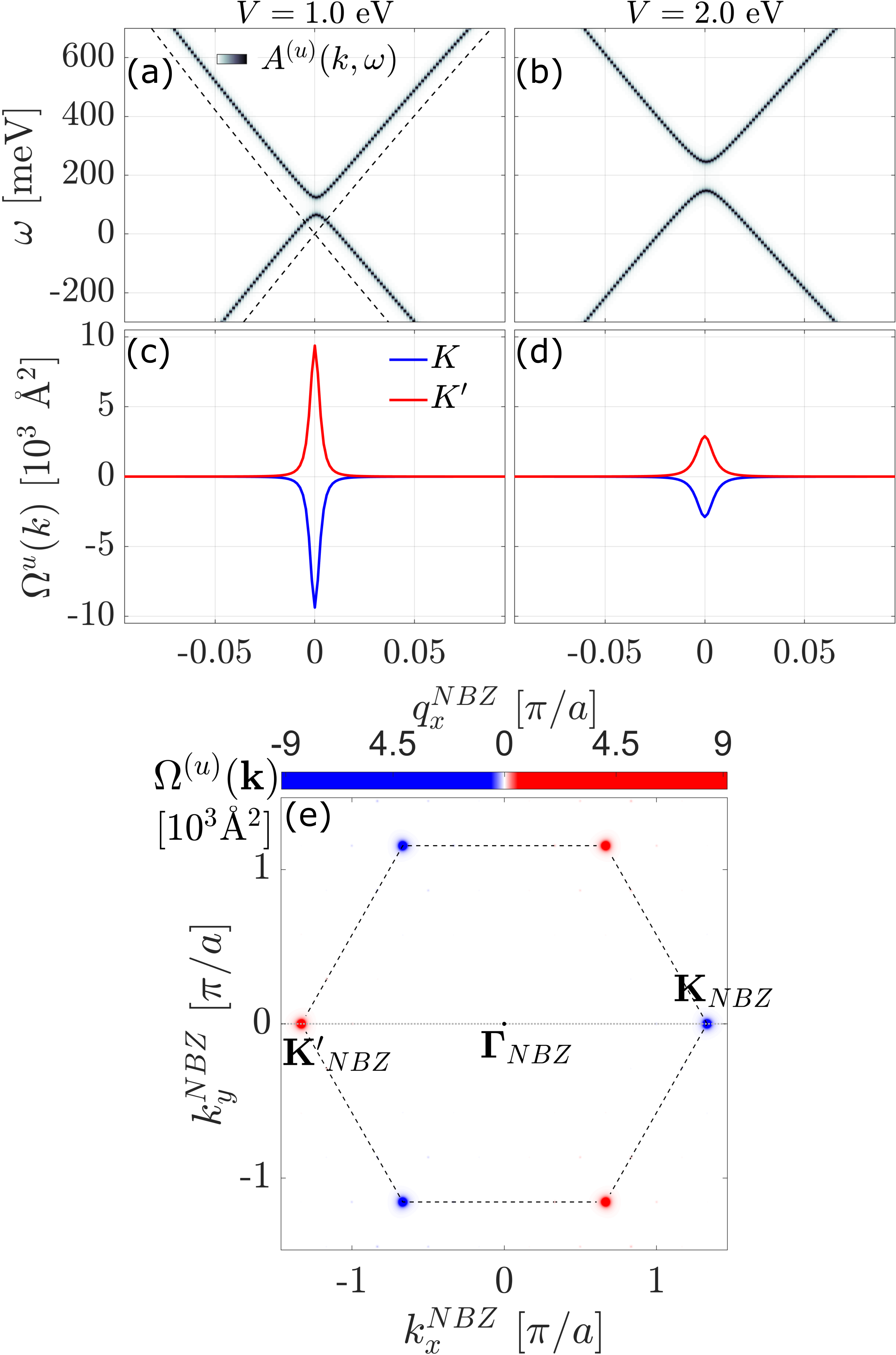}
\caption{
\fl{a}-\fl{b} Line-cuts of the unfolded spectral weight (gray surface) close to the NBZ $K$ point for different values of the constant superlattice potential $V(\rv_i) = 1$ eV, $2$ eV. The result at the $K'$ point along this same cut in $\kv$-space can be found by reflection around the central point $K_\tau$, and thus has similar structure.  
\fl{c}-\fl{d} Corresponding line-cuts of the unfolded occupied Berry curvature in the $K$ (blue) and $K'$ (red) valley  with the Fermi energy fixed in the gap at each potential. 
\fl{e} Unfolded Berry curvature in the NBZ demonstrating equal peaks of opposing signs, indicating the presence of transverse valley currents. The dotted line indicates the cut in $\kv$-space shown above.  
}
\label{fig:linecuts_UNF}
\end{figure}

We first consider the electronic structure of the superlattice of Fig. \ref{fig:intro}\fl{a} ($L = 4, R = 3$) directly in the SBZ. For $V = 0$ we recover the usual graphene band structure folded into the superlattice Brillouin zone [dashed lines in Fig. \ref{fig:linecuts_supercell}\fl{a}]. For the geometry considered here the $K, K'$ points are both folded in to the superlattice $\Gamma_{SC}$ point, resulting in nearly degenerate linear bands around this symmetry point. The splitting of these curves at larger $|k_{x}^{SBZ}|$ depends on the choice of the specific cut in $\kv$-space.  
When the finite superlattice potential is applied, an effective sublattice asymmetry is obtained on top of a constant overall shift of the bands. Thus, for $V \neq 0$ a gap opens continuously in the spectrum, with a simultaneous shift of the bands upwards in energy as shown in Fig. \ref{fig:linecuts_supercell}\fl{a}-\fl{b}. For the structures considered in this work the sublattice asymmetry is an intrinsic feature which is not removed by smoothly varying gate potentials, and we thus find these band gaps to be stable with respect to the smoothness of the applied potential with only a minor decrease in the gap magnitude (see Sec. III D. below). We note that the gap may close at larger values of $|V| \sim t$ depending on the specific geometry of the gated region and supercell width, but the gap formation at $|V| < t$ considered here is universal to all geometries, as predicted previously for potentials of $C_3$ symmetry. \cite{Malterre2011}  We demonstrate this universal gap formation in Figure \ref{fig:gap_shift_4}\fl{a} where the density of states (DOS) is shown for different extents of the superlattice potential in the supercell. The inset shows the corresponding band gap size ($\Delta$), and the shift in the center of the band gap ($E_s$) as a function of the superlattice potential magnitude $V$. The effect of varying the magnitude of the superlattice potential is similar to that of changing the ratio between the gated region (triangle side length $R$) and the supercell size (hexagon side length $L$), as investigated further in Appendix \ref{app:gap_shift}.
Similar gap openings have been demonstrated previously within the tight-binding model for gated superlattices in Ref. \onlinecite{Pedersen2012}, where circular potentials were considered instead. However, the gap opening in Ref. \onlinecite{Pedersen2012} was attributed to the local sublattice asymmetry near the edge, and thus these band gaps were found to be quickly decaying with increasing smoothness of the gate potential due to the disappearance of the local edge asymmetry. 

In Fig. \ref{fig:linecuts_supercell}\fl{c}-\fl{d} we show the supercell Berry curvature along the same cut in $\kv$-space as in \fl{a}-\fl{b}. The distribution displays a double peaked structure, with a clear sign change appearing exactly at the $\Gamma_{SC}$ point. As the superlattice potential is increased, this distribution is noticeably broadened but retains its shape. A similar result is obtained if the supercell and potential geometries are changed instead as shown in Fig. \ref{fig:gap_shift_4}\fl{b}. The full threefold symmetry of this distribution arising from the supercell folding is shown in Fig. \ref{fig:linecuts_supercell}\fl{e} where the Berry curvature is shown in the full SBZ. The rotational symmetry of this distribution follows from the specific folding of the NBZ valleys into the SBZ. The same symmetrical distribution is found when other superlattice geometries are considered, the only variation being in the width of the Berry curvature peaks. This effect is illustrated in Fig. \ref{fig:gap_shift_4}\fl{b}.  
\subsection{Unfolded Berry curvature and valley Hall conductivity}
Prior to our consideration of the unfolded result, it is instructive to compare the superlattice tight-binding calculations with results from a well-known model of the valley Hall effect in graphene. For this purpose, we consider a model which neglects confinement due to the periodic structure of the applied potentials, and simply considers the average potential on the A and B sites of the graphene system, leading to an effective sublattice asymmetry. This corresponds to a gapped Dirac model,
\begin{align}
	H_{\tau}(\vec{q}) &= \frac{\sqrt{3}}{2} at (\tau q_x \sigma_x + q_y \sigma_y )  + \frac{\Delta}{2} \sigma_z,  
\end{align}
with $\tau = \pm 1$ the valley index, $\vec{q} = \vec{k} -\tau \vec{K}$ measured with respect to the $K, K'$ points, and $a$ the graphene lattice constant.    
The Berry curvature in the $K, K'$ region close to the gap edge can be derived analytically, e.g. for the conduction band \cite{Xiao2007}
\begin{align}
	\Omega_{xy}(\vec{q}) &= \tau \frac{3 a^2 t^2 \Delta}{2(\Delta^2 + 3a^2 t^2 \abs{\vec{q}}^2)^{3/2}}, \label{eq:BC_an}
\end{align}
with associated Berry phases approaching $\pm \pi$ for small $\Delta$, and hence a quantized valley Hall conductivity following from Eq. (\ref{eq:VRHC}-\ref{eq:VHC}) of $\sigma^{v}_{xy} = 2e^2/h $ at the top of the valence band. This simple model with Berry curvature peaks of opposite sign in each valley and quantized valley Hall conductivity will serve as the comparison point for the superlattice results. We note that the utilized full tight-binding model goes beyond the simple decomposition into distinct valleys in the massive Dirac model above, since the tight-binding model includes both valleys and thus the effects of intervalley scattering. \cite{Cresti2016}

We now turn to the unfolded quantities $A^{(u)}, \Omega^{(u)}$, which are shown in Fig. \ref{fig:linecuts_UNF}. The spectral weight of the nearly degenerate bands in the supercell around the $\Gamma_{SC}$ point now unfold into the NBZ $K, K'$ valleys as seen from the line-cut through the $K$ point in \fl{a}-\fl{b}. As such, the unfolded spectral weight resembles the valley structure of the massive Dirac model introduced above. Correspondingly, the unfolded Berry curvature peaks exactly at the center of each valley, but with opposite signs as shown in Fig. \ref{fig:linecuts_UNF}\fl{c}-\fl{d}. The full distribution is shown in Fig. \ref{fig:linecuts_UNF}\fl{e}. Here, we observe sharp peaks around each symmetry point with opposite signs in the entire valley regions. It now becomes clear how the rotational symmetry of the supercell Berry curvature arises. The unfolded Berry curvature peaks of each valley fold into separate regions of the SBZ around the $\Gamma_{SC}$ point, yielding the flower structure in Fig. \ref{fig:linecuts_supercell}\fl{e}. 

\begin{figure}[tb]
\includegraphics[width = \linewidth ]{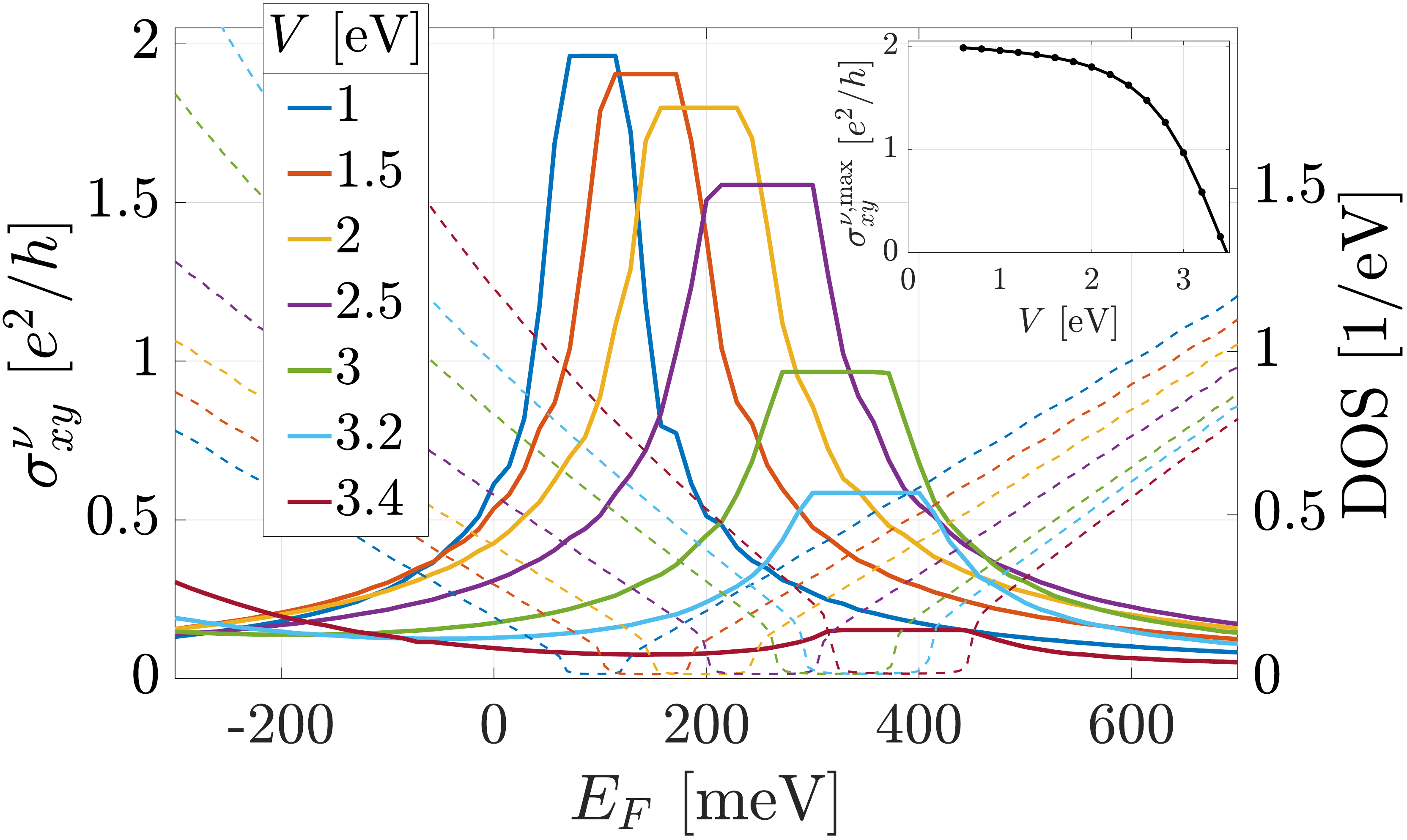} 

\caption{
 Valley Hall conductivity as a function of filling (full lines) for varying values of the superlattice potential $V$, shown alongside the density of states (dotted lines). Berry curvature accumulated near the band edges causes a saturation of the valley Hall conductivity as the gap is approached, and for small $V$ the quantized $2e^2/h$ value of the massive Dirac model is approached. The inset shows the plateau value in the gap as the superlattice potential is tuned. The valley Hall conductivity decays for larger superlattice potentials, as the supercell bands flatten and the unfolded valley structure is lost. 
 }
\label{fig:sigma_flat}
\end{figure}

A finite valley Hall effect in these systems is evident from the unfolded Berry curvature distribution, since integration of this quantity around each valley yields finite valley-resolved conductivities of opposite signs. The result of the integration procedure [Eq. (\ref{eq:VRHC}-\ref{eq:VHC})] is shown in Fig. \ref{fig:sigma_flat} as a function of the Fermi energy for different values of the superlattice potential. As demonstrated above, the band edges act as Berry curvature hot spots causing a saturation of the valley Hall conductivity as the Fermi energy approaches the gap from below. This plateau then decays when states in the bands above the gap start contributing Berry curvature of opposite sign. In the limit of small $V$ we found above that the unfolded electronic structure and Berry curvature distribution closely resembles an effective massive Dirac model, and in this case we also find that the valley Hall conductivity approaches a quantized plateau value of $2e^2/h$ as predicted from Eq. (\ref{eq:BC_an}). When the superlattice potential is increased this plateau widens as the gap expands and a small variation in the plateau value appears. We note that the numbers of k points needed to converge the valley Hall conductivity increase dramatically as the potential is decreased since the Berry curvature distribution becomes more sharply peaked. All calculations in this work are performed with $N_{\kv} = 230 \times 230$ $\kv$-points.   

In the limit of larger superlattice potentials the simple resemblance to the shifted massive Dirac model breaks down, and the valley Hall conductivity decays from the quantized plateau value of $2e^2/h$ as demonstrated in Fig. \ref{fig:sigma_flat}, ultimately vanishing at $V = 3.4$ eV. 
In this limit the superlattice potential approaches the energy scale of the hopping $t$ and the electronic structure is strongly perturbed, resulting in a Berry curvature distribution diverging from the simple model. In particular, the valence and conduction bands flatten and the valley-structure of the unfolded spectral weight is lost.

\begin{figure}[tb]
\includegraphics[width = \linewidth]{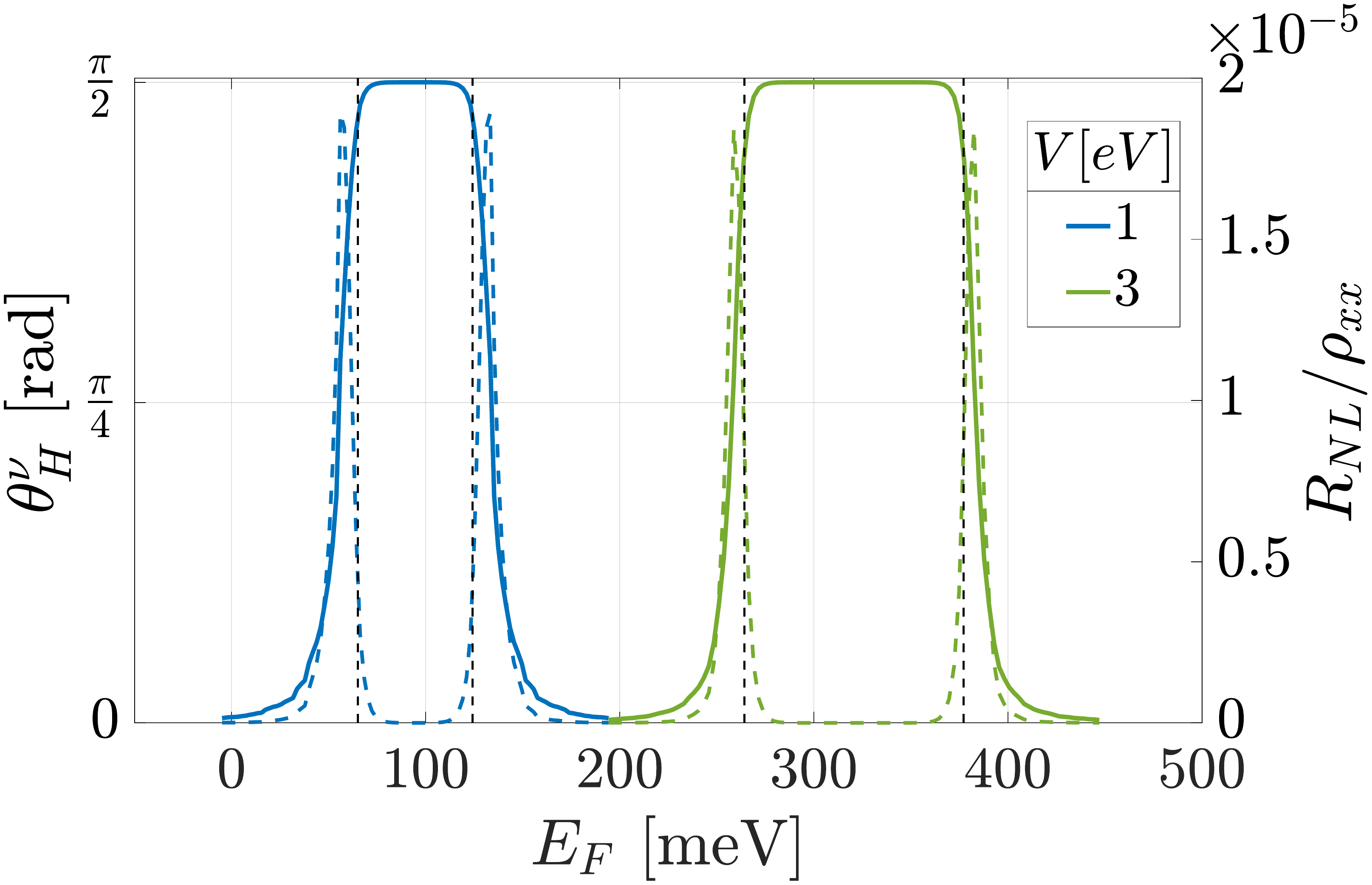}
\caption{
	 Valley Hall angle (full lines) and expected nonlocal resistance signal (dashed lines) close to the band edge for two values of the superlattice potential $V = 1$ eV, $3$ eV. The band gap is indicated by the vertical dashed lines. The valley Hall angle is only finite close to the band edge where $\sigma_{xy} \sim \sigma_{xx}$, and approaches $\pi/2$ in the gap.  The predicted nonlocal resistance close to the band edges is obtained using the expression of Ref. \onlinecite{Beconcini2016}. The peaks in the ratio $R_{NL}/\rho_{xx}$ occur exactly at the $\theta_{v} = \pi/4$ point, i.e. when the valley Hall and longitudinal conductivities are equal, $\sigma_{xy}^{v} = \sigma_{xx}$. These peaks in the nonlocal response shift as the superlattice potential is tuned.   
}
\label{fig:nonlocal_res}
\end{figure}

\subsection{Valley Hall angle and associated nonlocal response}
In Fig. \ref{fig:nonlocal_res}\fl{a} we show the valley Hall angle $\theta_H^v  = \arctan{\sigma_{xy}^v/\sigma_{xx}}$, which is the ratio of the magnitude of the transverse valley and longitudinal charge currents. The angle is finite only close to the band edge where the valley Hall conductivity peaks and exceeds the longitudinal conductivity in a small interval.
Following Ref. \onlinecite{Beconcini2016}, we estimate the valley Hall contribution to the nonlocal resistance from the valley Hall angle in, e.g., a Hall bar of width $W$, with inter-terminal distance $d$, and valley diffusion length $l_v$:
\begin{align}
\Delta R_{NL}/\rho_{xx}  &= \frac{W}{2L_v}\frac{\tan^2 \theta_{v}}{1 + \tan^2 \theta_{v}} e^{-\abs{d}/L_{v}}, 
\end{align}
where $L_{v} = l_{v} \sqrt{1+\tan^2 \theta_{v}} $ is a renormalized valley diffusion length. 
 
We note that this interpretation relies on the picture of bulk valley currents carried by subgap states,\cite{Lensky2015,Beconcini2016} which is but one interpretation of nonlocal measurements in valley Hall systems. In particular, these currents are missing when the Fermi energy is placed in the gap in Landauer-B{\"u}ttiker calculations, \cite{Kirczenow2015} and only reappear as edge currents when detailed modeling of the electronic structure and edge profiles are considered. \cite{Marmolejo_Tejada2017} In this work we thus restrict ourselves to making predictions close to the band edge outside the gapped region where the interpretation as bulk valley currents is valid. 

The expected nonlocal signal for varying values of the superlattice potential is displayed in Fig. \ref{fig:nonlocal_res}\fl{b}, for $W, d, l_v = 100, 10^3, 10^5$ nm. The nonlocal response is shifted as the the superlattice potential is varied, since it peaks near the band edge where the valley Hall angle $\theta_{v}$ approaches $\pi/4$. This tunability of the nonlocal response with the external potential provides an unambiguous way of separating stray current and valley Hall contributions to the nonlocal resistance.

\subsection{Robustness with respect to the dielectric environment}

\begin{figure}[tb]
\includegraphics[width = \linewidth ]{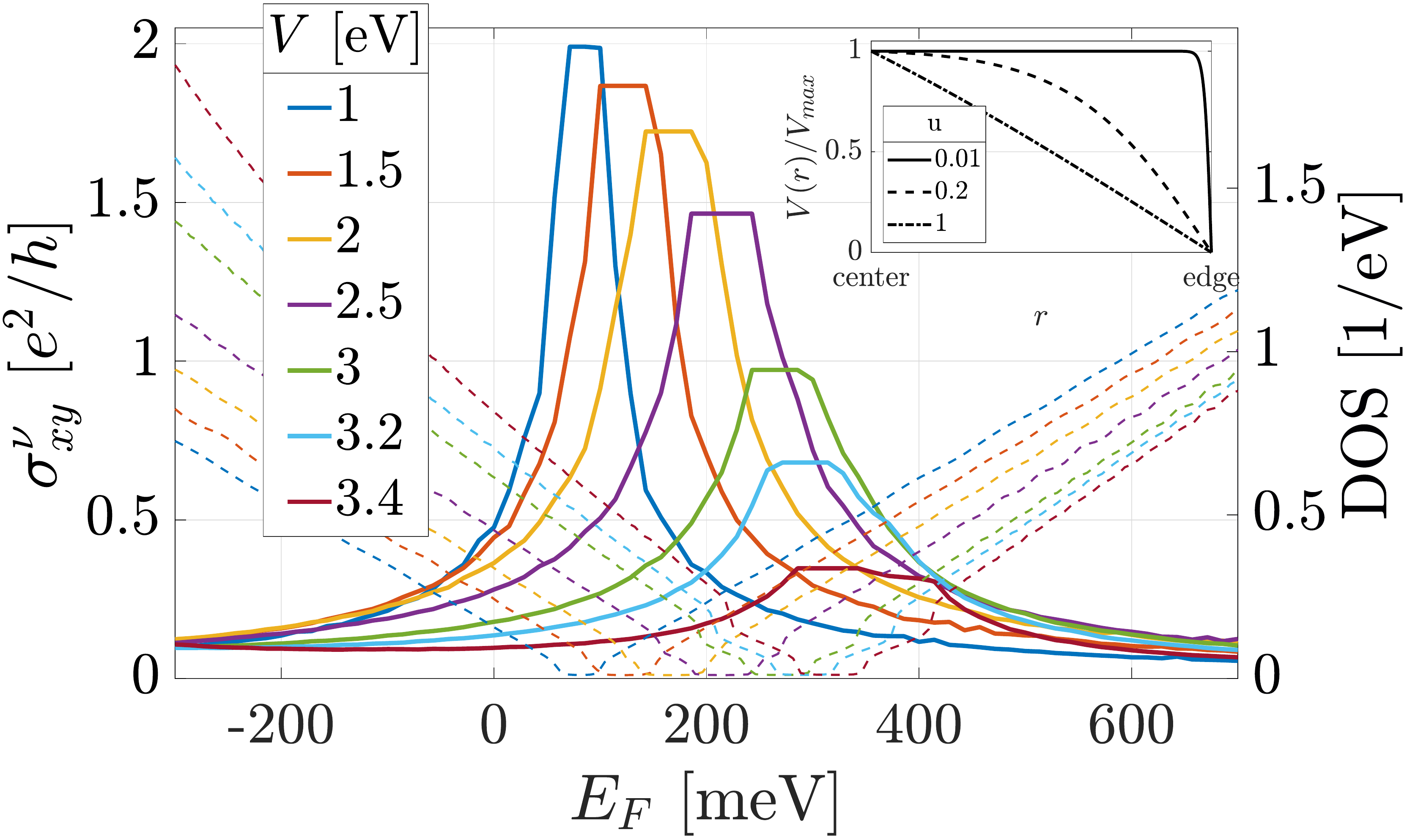}
\caption{
 Valley Hall conductivity as a function of filling for different values of the superlattice potential for a smoothly varying potential ($u = 0.2$), the profile of which is displayed in the inset. The results are similar to the flat potential case, with some additional structure in the peak structure due to the lifting of degeneracies of bands near the band edge. 
 }
\label{fig:sigma_smooth}
\end{figure}

In what follows we consider more realistic potentials based on the specific dielectric environment in patterned dielectric superlattices. In particular, we consider potentials varying smoothly with the distance $r$ from the edge of the side of the nanostructured indentation in the dielectric to the center, here parametrized by  
$V(r)/V_{max} = [\exp((r-1)/u)+1]^{-1}-1/2$, 
with $u \in [0,1]$ a continuous parameter setting the smoothness of the potential, $u = 0$ being the flat potential considered so far, and $u = 1$ the extreme case of a linearly decreasing potential. Line profiles of this potential are shown in the inset of Fig. \ref{fig:sigma_smooth}, and the full 2D potential for $u = 0.2$ is shown in the gradient of Fig. \ref{fig:intro}\fl{a}. Further details of the spatial profile of the smoothly varying potential are included in Appendix \ref{app:potential}. 

The valley Hall conductivity obtained for this potential is shown in Fig. \ref{fig:sigma_smooth}. The result is similar to that obtained above for the flat potential, although with slightly narrower plateau regions. Additionally, new features appear away from the band edge since degeneracies are lifted and thus the integrated Berry curvature varies in small increments when each band edge is reached. For small potentials we again approach the quantized value in the gap. 

\begin{figure}[tb]
\includegraphics[width = \linewidth ]{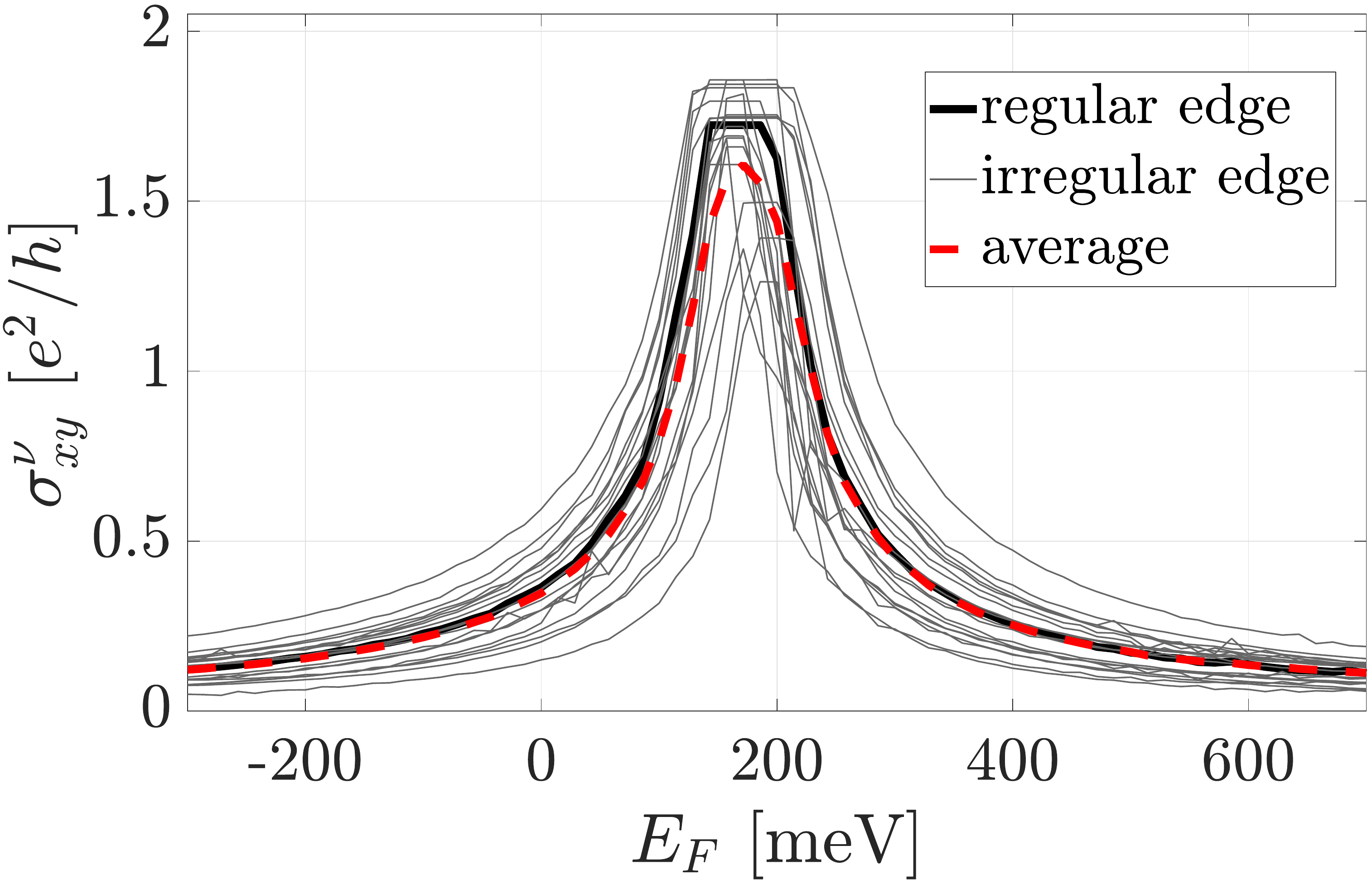}
\caption{
Variation of the valley Hall conductivity with respect to irregularities in the edge profile of the superlattice potential, corresponding to irregularities in the dielectric etching. The regular limit  for a smoothly varying potential ($V = 2$ eV, $u = 0.2$) is shown in the full black line, alongside the same calculation with random edge profiles at the superlattice potential boundary (gray lines). The average of all such configurations is shown in the red dotted line. The finite valley Hall conductivity does not require a perfectly symmetrical induced potential, and is thus a general prediction in these superlattices.  
 }
\label{fig:stability}
\end{figure}

Finally, we conclude our analysis of realistic potentials by considering irregularities in the edge of the dielectric etching, which modulates the potential near the edge. We simulate this effect by adding a random potential to the edges of the gated region in the supercell, disrupting the perfect zigzag edges considered thus far which where expected to minimize coupling of the valleys. At each site which is a nearest neighbor to the gate regi
on edge we add a random potential $w_{edge} \in [-0.5,0.5]$ eV, and consider the resulting valley Hall conductivity for different random configurations at a fixed superlattice potential ($V = 2$ eV, $u = 0.2$). The result of this procedure is shown in Fig. \ref{fig:stability} (gray lines), together with the clean limit result (full black line), and the average of the irregular configurations (red dashed line). The application of these random edge potentials does not substantially modify the valley Hall conductivity, which displays a shifted peak structure for all configurations with a small variation in the plateau value. The average tracks the clean result peak, with a rounded plateau due to the different shifts of the gapped region in different configurations. 

\section{Discussion and conclusions}

We have theoretically investigated graphene superlattices defined by periodic gating as a platform for valleytronics. For zigzag edged triangular potentials where inversion symmetry is broken and intervalley scattering is suppressed, a gate-tunable valley Hall effect appears. This effect stems from the accumulation of Berry curvature near the band edge of the superlattice band structure, which unfolds to curvature of opposite sign in the $K$ and $K'$ valleys of the graphene Brillouin zone. For small potentials the system resembles a gapped Dirac model with quantized valley Hall conductivity, yet when the gate-tunable potential is increased this valley Hall conductivity decreases continuously, resulting in a platform for valleytronics where both the magnitude and width of the valley Hall conductivity plateau can be tuned by an external gate. Finally, we have considered experimental signatures of the gate-tunable valley Hall effect when the Fermi energy is tuned close to the band edge in nonlocal transport experiments, and determined how this response varies with the external potential.  

In this work we have considered the maximum of the externally induced potential as the tunable parameter. In addition to this degree of freedom the effect of alignment between the substrate and the graphene sheet, with a corresponding rotation and shift in the induced potential, can also have a profound impact on the valley Hall conductivity. \cite{Jung2018} For the atomically-resolved model considered here the result will in general depend on the size of the gated region, with sign changes in the valley Hall conductivity when the sublattice is shifted. 

Our idealized model of irregularities at the edge of the induced potential implies a periodic structure with the same edge profile, and as such we are limited to calculating modifications to the intrinsic part of the valley Hall conductivity. In general the valley Hall conductivity also has contributions from disorder, commonly classified as the side-jump and skew scattering corrections. \cite{Sinitsyn2007} We note that these corrections occur outside the gapped region, and do not substantially modify tunable properties of the valley Hall conductivity in these systems. \cite{Ando2018}  

The main measurable consequence of the nonzero Berry curvature in time-reversal invariant systems, such as the superlattice considered in this work, is a finite correction to the nonlocal resistance. Recently, additional measurable consequences have been predicted, including applications in current rectification, \cite{Isobe2018} and direct detection via the so-called Magnus Hall effect. \cite{Papaj2019} The gate-tunable Berry curvature predicted in this work could define a controllable platform for further investigations of these effects. 

\section{Acknowledgments}
We acknowledge useful discussions with S. Power. The Center for Nanostructured Graphene is supported by the Danish National Research Foundation, Project DNRF103.

\FloatBarrier
\appendix

\section{Unfolding procedure} \label{app:UNF}
We unfold quantities calculated in the supercell Brillouin zone (SBZ) back into the pristine graphene, or normal, Brillouin zone (NBZ) following Ref. \onlinecite{Ku2010}. 

Real space and reciprocal lattice vectors in the normal- and supercell are related by \cite{Popescu2012}
\begin{align}
\vec{A} &= \mat{M} \cdot \vec{a},\\
\vec{B} &= \mat{M}^{-1} \cdot \vec{b}, \label{eq:vec_transform}
\end{align}
with $\mat{M}$ a matrix of integers. \\
For the triangular superlattices considered here, the general form of this matrix is \cite{Guinea2010}
\begin{align}
\mat{M} &= L\begin{pmatrix}
2 & 1 \\
1 & 2 
\end{pmatrix}, 
\end{align}

with $L$ the side length of the supercell hexagon. The determinant of this matrix is the ratio of unit cell volumes. \\ 
A given wavevector $\kv \in$ NBZ is folded into a unique $\Kv \in$ SBZ by a reciprocal lattice vector \cite{Popescu2012}
\begin{align}
\vec{K} &= \vec{k} - \vec{G}_0,   \label{eq:k_fold}
\end{align}
with $G_0 = \sum_i q_i \vec{B}_i$, where the $q_i$ are integers. 
We define $\Kv'(\kv)$ as the unique $\Kv$ point to which a given $\kv$ point folds. \\
A wavevector in the SBZ unfolds into multiple values 
\begin{align}
\vec{k}_i &= \vec{K} + \vec{G}_i,  
\end{align}
with a number of elements $N_k$ in $\{G_i\}$ given by $N_k = \det \mat{M}$. \cite{Popescu2012} \\

We employ a tight-binding calculation using localized orbitals $\ket{\phi_{i\rv}}$, and find the Bloch states. These are characterized by quantum number $n$ and wavevector $\kv$ in the normal (pristine) cell, and by quantum number $N$ and wavevector $\Kv$ in the supercell
\begin{align}
\ket{n\kv} &= \sum_{i} C_{in\kv}  \ket{i\kv} \\
&= \sum_{i\rv} C_{in\kv}  e^{i\kv \cdot (\vec{r}+\tauv_i)} \ket{\phi_{i\rv}}, \\
\ket{N\Kv} &= \sum_{I\Rv} C_{IN\Kv} e^{i\Kv \cdot (\vec{R}+\tauv_I)} \ket{\phi_{I\Rv}},
\end{align}
with $\rv, \Rv$ lattice vectors in the normal cell and supercell, and $\tauv_{i/I}$ the relative position of each orbital in the unit cell and supercell, respectively.\\

Given an quantity $\mathcal{O}_{N\vec{K}}$ defined in the SBZ, we now define the corresponding unfolded quantity in the NBZ:
 \begin{align}
 \mathcal{O}_{i\vec{k}}^{(u)} &= \sum_{NK} \abs{\braket{i\vec{k}}{N\vec{K}}}^2 \mathcal{O}_{N\vec{K}} \label{eq:unfold} \\
 &=  \sum_{N} \lambda_{iN\kv} \mathcal{O}_{N\Kv'(\kv)} \label{eq:O_unfolded}. 
 \end{align}
Unfolding then boils down to finding the Bloch state overlap $\lambda_{iN\kv}$, which we will derive within a tight-binding scheme below. Note that the unfolding becomes more complicated for the Berry curvature since a derivative with respect to $\kv$ is included in the definition of this quantity (see Ref. \onlinecite{Bianco2014} eq. 31).

Define a map $I \to \Rv+\rv'(I), i'(I)$ uniquely identifying a localized orbital in the supercell ($I$) with a similar orbital in the normal cell [$i'(I)$], where $\rv'(I)$ is a normal cell lattice vector giving the relative position between unit cells. We can then calculate the overlap between a given supercell and normal cell orbital:
\begin{align}
\braket{\phi_{i\rv} }{\phi_{I\Rv} } &= \braket{\phi_{i\rv} }{\phi_{i'(I)\Rv+\rv'(I)} }  \\
&= \delta_{ii'(I)} \delta_{r,\Rv+\rv'(I)},
\end{align} 
where the final equality follows from orthogonality of the normal cell orbitals. This simple form of the orbital overlap enables a calculation the Bloch state overlap

\begin{align}
 \lambda_{iN\kv} &=  \braket{i\kv}{N\Kv} \\
&= \sum_{I,\rv \Rv} C_{IN\vec{K}} 
	e^{-i\vec{k}\cdot (\rv+\tauv_i)}e^{i\vec{K}\cdot (\vec{R}+\tauv_I)}  \braket{\phi_{i\rv} }{\phi_{I\Rv} } \\
&= \sum_{I,\Rv} C_{IN\vec{K}} 
	e^{-i\vec{k} \cdot (\Rv+\rv'(I)+\tauv_i)}e^{i\vec{K}\cdot (\vec{R}+\tauv_I)} \delta_{ii'(I)} \\
&= \sum_I C_{IN\Kv} e^{-i\kv \cdot (\rv'(I)+\tauv_{i})} e^{i\Kv \cdot \tauv_I} \delta_{ii'(I)} \delta_{\Kv[\kv]},
\end{align}
where $[\kv]$ is the set of wavevectors $\kv$ which downfold to $\Kv$. Note that for a given $\kv$ 
the value of $\Kv$ for which this delta function is finite is unique. This enables us to collapse all sums over $\Kv$ when unfolding, picking out the value $\Kv'(\kv)$.\\

Calculation of the unfolded Berry curvature proceeds from this formalism using the gauge-invariant approach of Ref. \onlinecite{Bianco2014}, and its extension to tight-binding in Ref. \onlinecite{Olsen2015}.

\section{Band gap and shift for different geometries} \label{app:gap_shift}

\begin{figure}[tb]
\includegraphics[width = \linewidth ]{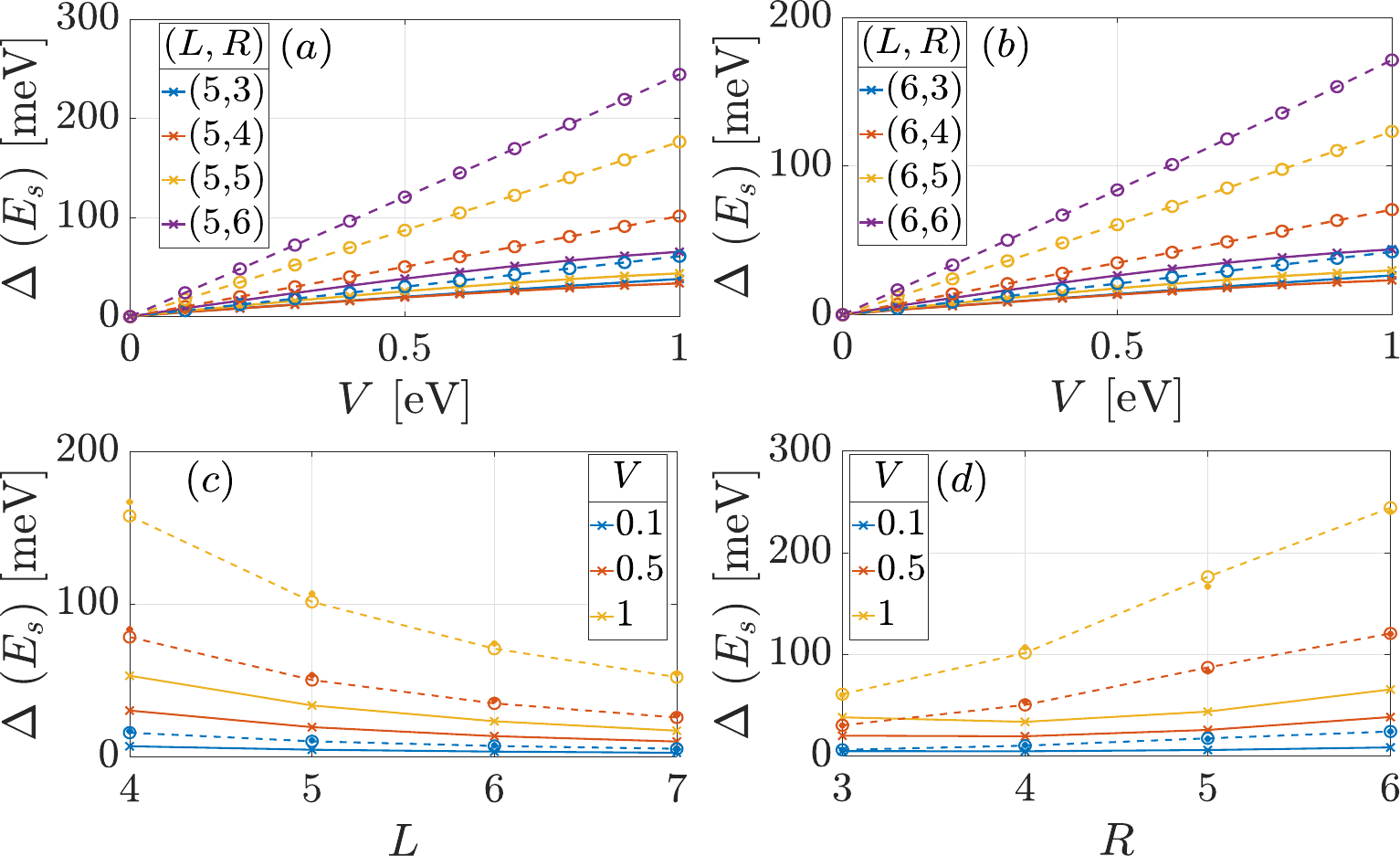}
\caption{\fl{a}-\fl{b} Band gap variation with the superlattice potential magnitude for different geometries. \fl{c}-\fl{d} Band gap ($\Delta$) and shift ($E_s$) variation with the supercell size $(L, R = 4)$ and extent of the superlattice potential $(L = 5, R)$, respectively, shown here for multiple values of the superlattice potential magnitude ($V$). The small asterisks indicate the average potential on each site in the supercell, which matches the numerically calculated shift ($E_s$). 
The same general result is obtained for different configurations: The band gap widens for either greater magnitude of the superlattice potential, or for increasing ratio between gated region and supercell size. }
\label{fig:gap}
\end{figure}

In this appendix we provide further information on the evolution of the gap in the spectrum $\Delta = \abs{E_{1} - E_{0}}$, and the shift in the center of this gap $E_s = E_{0} + \Delta/2$, where $E_{1, 0}$ indicate the band edges with $E_{1} > E_{0}$. Fig. \ref{fig:gap} displays further calculations of these quantities for different geometries \fl{a}-\fl{b}, and their evolution with the superlattice geometry parameters $L,R$ \fl{c}-\fl{d}. The shift in the center of the gap ($E_s$) is seen to vary linearly with the superlattice potential, as might be expected from considering the average potential in the unit cell. Indeed, calculating this average potential as $V_{avg} = V~(N_V / N_{SC}) \propto (R^2 / L^2)$, where $N_V = R^2$ is the number of sites with shifted onsite potentials from the superlattice potential and $N_{SC} = 6 L^2$ is the total number of sites in the supercell, we find a close match with the obtained value of of the shift. This average potential is shown (small asterisks) alongside the obtained shifts in Fig. \ref{fig:gap} \fl{c}-\fl{d}. 

Similar simple models for the band gap ($\Delta$) in the electronic spectrum of a given geometry based on, e.g., the average graphene A/B site asymmetry do not match the calculated band gap in these systems. This follows from the fact that band gap formation can be driven both by the periodic structure of the superlattice potential itself, which can result in band gaps even for circular potentials, and effects associated with local symmetry of the potential structure such as A/B asymmetry on the edges of the potential. The former of these mechanism can yield extreme sensitivity to small variations in the superlattice size $L$, as seen in, e.g., antidot lattices. \cite{Pedersen2008} For the potentials of $C_3$ symmetry considered in this work we thus restrict ourselves to the general observations, as found in similar superlattices, \cite{Pedersen2008, Pedersen2012} that the size of the induced band gap is directly proportional to the magnitude of the superlattice potential and the extent of this potential $R$, and inversely proportional to the supercell size ($L$), i.e., the distance between gated regions, as demonstrated in Fig. \ref{fig:gap}. 

\begin{figure}[tb]
\includegraphics[width = \linewidth ]{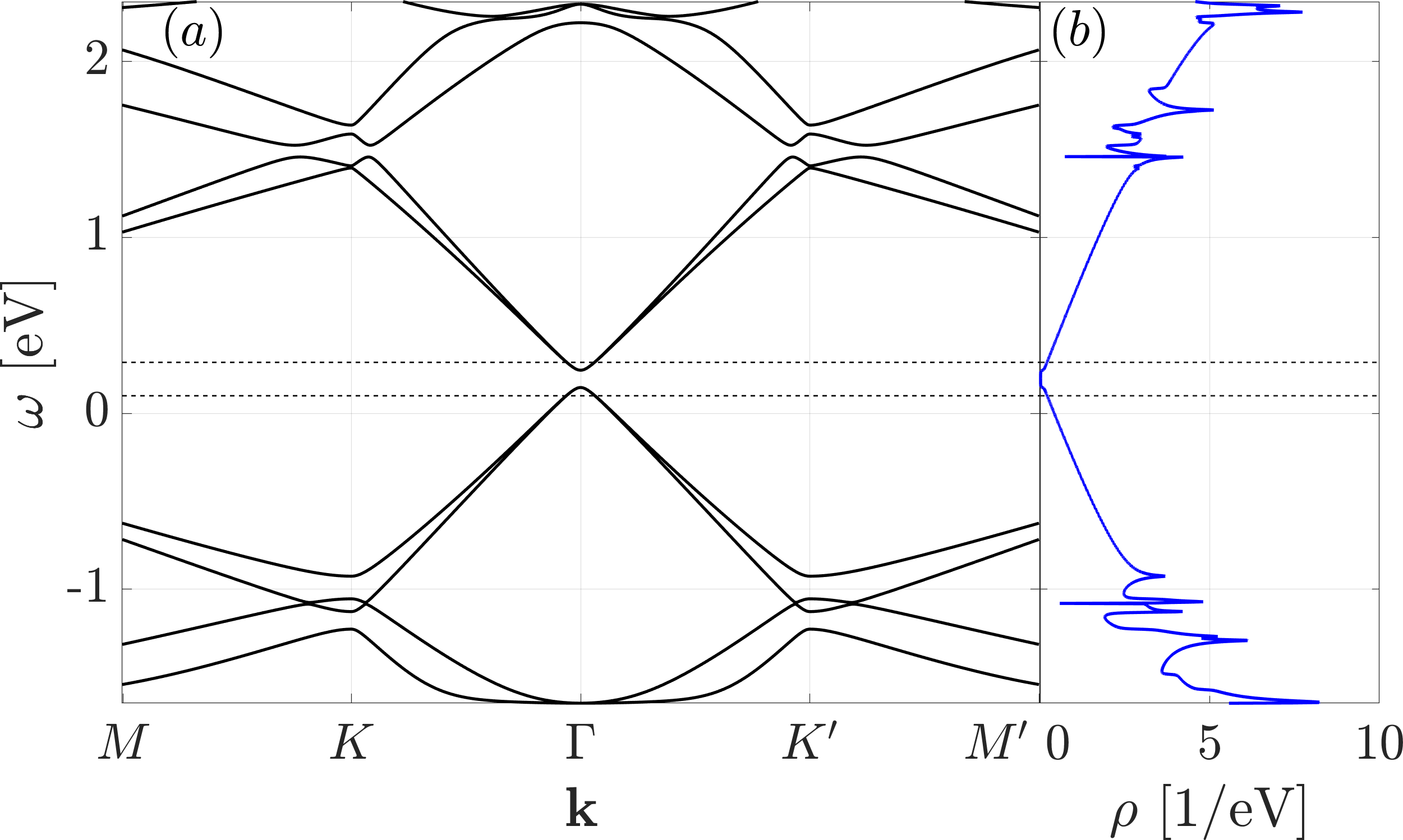}
\includegraphics[width = \linewidth ]{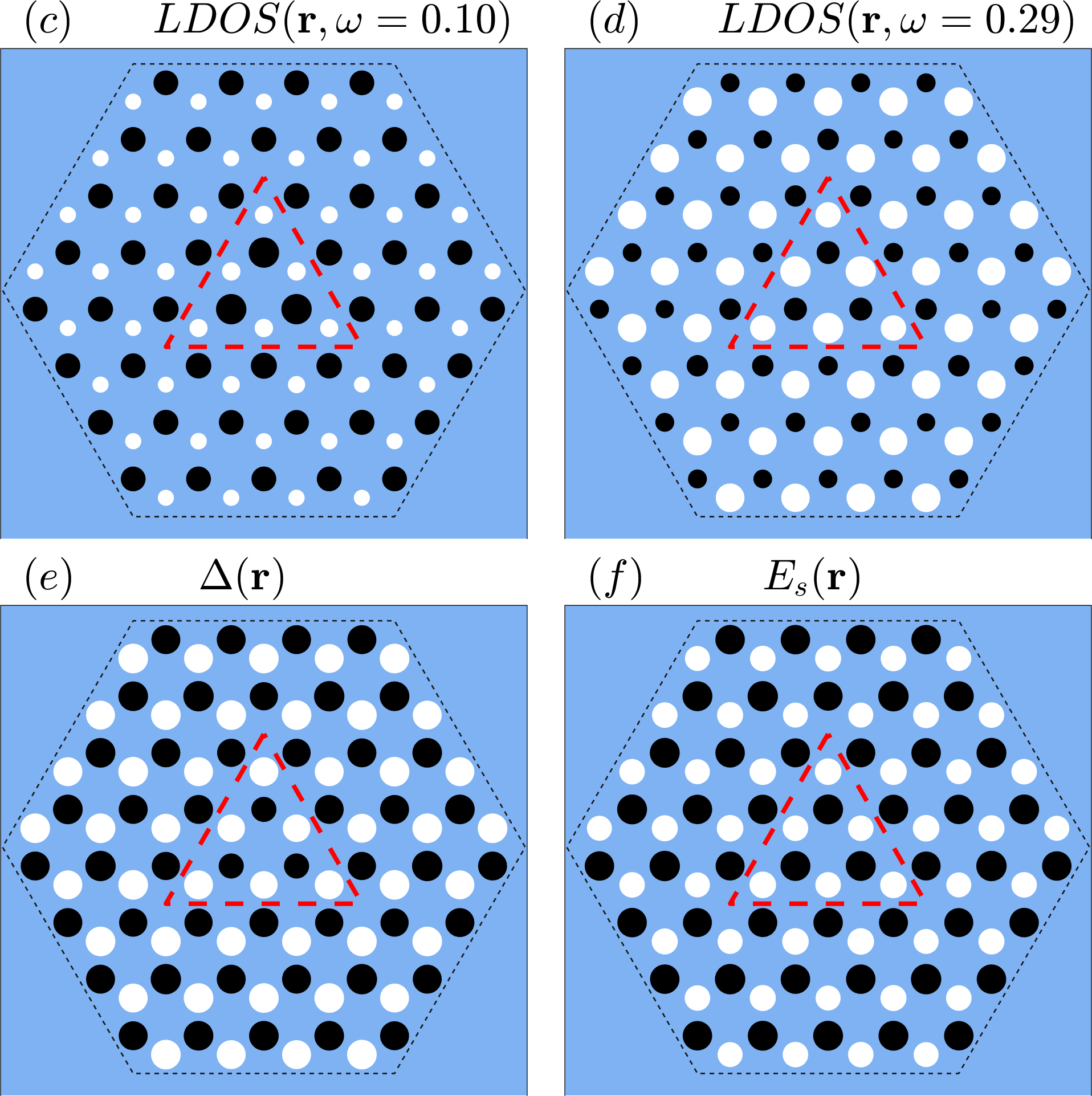}
\caption{\fl{a} Extended view of the band structure $E_N(\kv)$ showing the gap and miniband formation of the supercell. The symmetry points are those of the SBZ. \fl{b} Corresponding density of states, demonstrating the shifted band gap. \fl{c}-\fl{d} LDOS, plotted using the radii of black (white) disks to indicate the value at A (B) sites, sampled just above and below the gap at $\omega = 0.1, 0.29$ eV (dashed lines in \fl{a}). The superlattice potential breaks inversion symmetry and causes a splitting of the A/B weight at these sites. \fl{e} Local gap magnitude at each site in the supercell as derived from the local density of states (variations enhanced $\times 5$), showing a small variation at the potential edge. \fl{f} Corresponding shift in the center of this local gap (variations enhanced $ \times 5$), showing a small difference between A/B sites in the supercell. All plots are for a representative configuration of ($L = 4$, $R = 3$, $V = 2$ eV). } 
\label{fig:more_gap}
\end{figure}

A full picture of a typical band gap and the minibands closest to the gap is provided in Fig. \ref{fig:more_gap} \fl{a}-\fl{b}, alongside the DOS in the same region. The gap formation in real space can be observed by calculating the local density of states (LDOS) as a projection of the spectral weight on a given orbital $\phi_{I\Rv}$ in the supercell
\begin{align}
LDOS(\Rv_I, \omega) &= \frac{1}{N_{\Kv}} \sum_{N\Kv} \abs{\braket{\phi_{I\Rv}}{N\Kv}}^2 A_{N\Kv}(\omega). 
\end{align}
The LDOS above and below the band edge is displayed in Fig. \ref{fig:more_gap} \fl{c}-\fl{d} at energies as shown by the dashed lines in \fl{a}, and demonstrates the opposite splitting of the LDOS on the $A/B$ sublattices above and below the band edge caused by the inversion-symmetry-breaking superlattice potential. In these plots the LDOS is plotted on A (B) sites as black (white) disks, with the radius indicating the magnitude of the LDOS normalized to the maximal value in the supercell. 
From the LDOS around the gap region we can define the local gap and shift ($\Delta(\rv), E_S(\rv)$) using the band edges of the local gap in the LDOS at a given site. These quantities are shown in Fig. \ref{fig:more_gap} \fl{e}-\fl{f}, using a similar plotting scheme similar to that for the LDOS. In these cases the maximal variation from the mean is much smaller ( $7\%, 6\%$ for the gap and shift, respectively) than for the LDOS, and we have thus enhanced the variation fivefold in these plots. The local gap is almost homogeneous, and the only variation of the local gap magnitude $\Delta(\rv)$ is seen to take place close to the superlattice potential edge where the potential locally breaks A/B symmetry. The shift $E_s(\rv)$ is also homogenous apart from a minor constant A/B variation due to the different number of A/B sites enclosed by the superlattice potential.

\section{Spatial profile of the superlattice potential} \label{app:potential}

\begin{figure}[tb]
\includegraphics[width = \linewidth ]{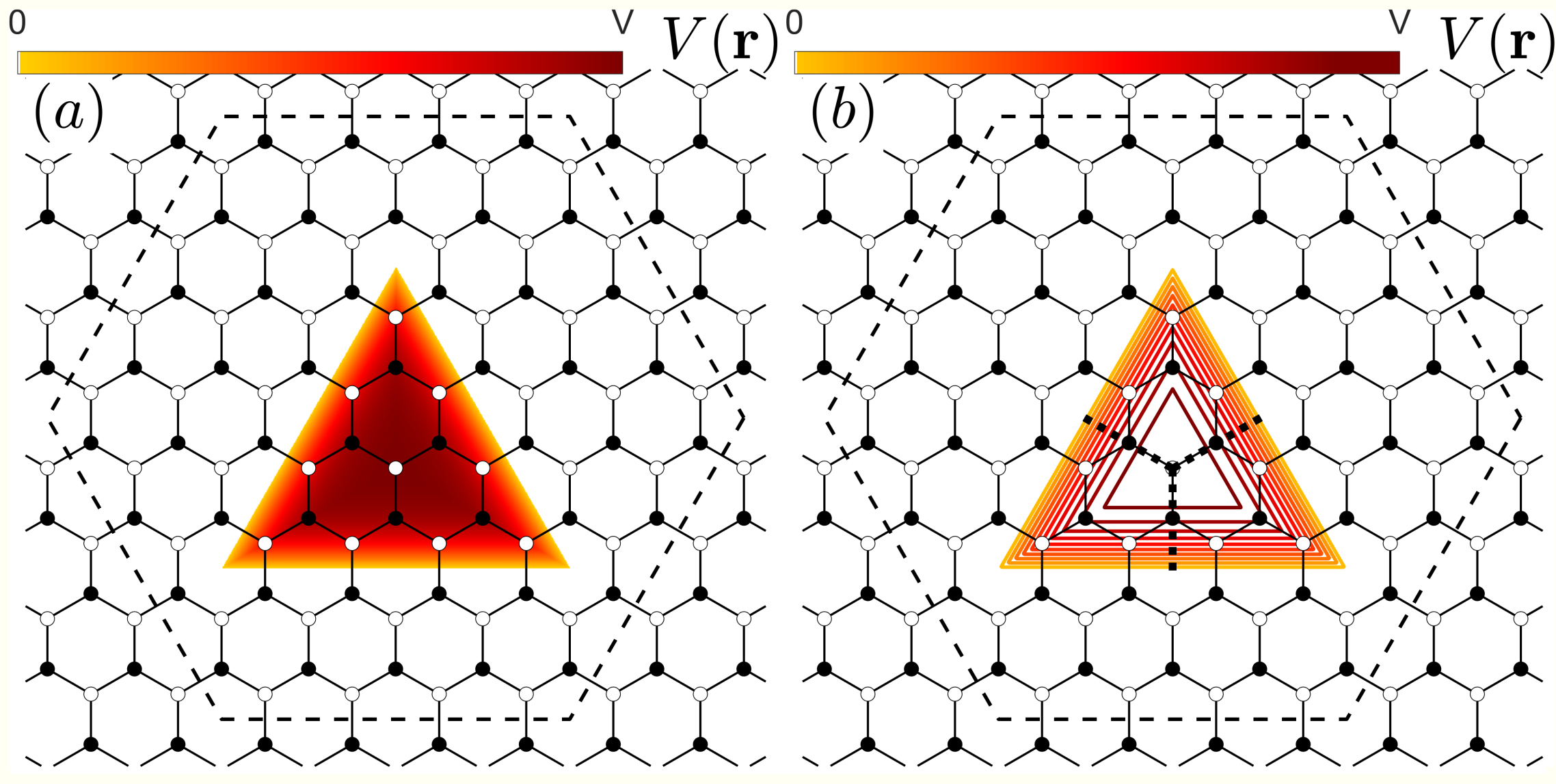}\\
\includegraphics[width = \linewidth ]{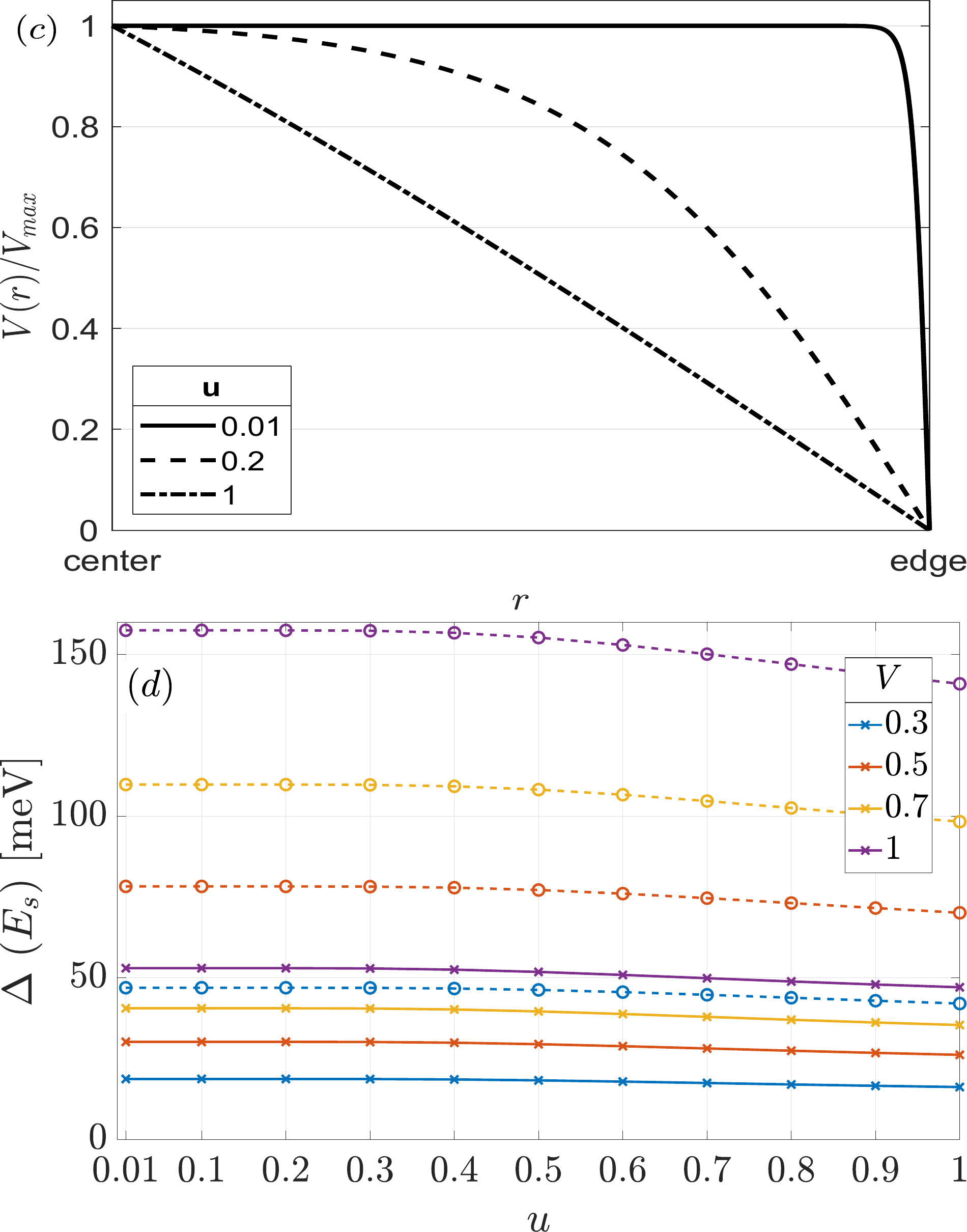}
\caption{Spatial variation of the superlattice potential ($[L, R, u] = [4, 4, 0.2]$), shown as \fl{a} a color gradient, \fl{b} a contour plot. \fl{c} The equivalent linecuts indicated by the black dotted lines in \fl{b} for different values of the smoothness parameter $u = [0.01, 0.2, 1]$, which interpolates between the extreme cases of flat and linearly decreasing potentials.  \fl{d} Variation of the induced band gap (full lines) and shift (dashed lines) with the smoothness parameter $u$ for the $(L, R) = (4,4)$ geometry outlined above. There is only a small decay in the gap magnitude.}
\end{figure}

\end{document}